\documentclass[sigconf]{acmart}

\usepackage{booktabs} 
\usepackage{tabularx}
\usepackage{multirow}
\usepackage{hyperref}

\setcopyright{none}
\setcopyright{rightsretained}

\usepackage[textsize=scriptsize,backgroundcolor=red!10,bordercolor=red!90]{todonotes}

\renewcommand\footnotetextcopyrightpermission[1]{}  

\acmDOI{}
\editor{Srishti}

\begin{document}
\title{Heterogeneous Edge Embeddings for Friend Recommendation}

\author{Janu Verma, Srishti Gupta, Debdoot Mukherjee}
\affiliation{%
  \institution{Hike Messenger, India}
 }
\email{{srishti, janu, debdoot} @hike.in}

\author{Tanmoy Chakraborty}
\affiliation{%
  \institution{IIIT Delhi, India}
 }
\email{tanmoy@iiitd.ac.in}


\begin{abstract}
 We propose a friend recommendation system (an application of link prediction) using edge embedding on social networks. Most real world social networks are multi-graphs, where different kinds of relationships ({\em e.g.,} chat, friendship) are possible between a pair of users. Existing network embedding techniques do not leverage signals from different edge types and thus perform inadequately on link prediction in such networks. We propose a method to mine network representation that effectively exploits edge heterogeneity in multi-graphs. We evaluate our model on a real-world, active social network where this system is deployed for friend recommendation for millions of users. Our method outperforms various state-of-the-art baselines 
on Hike's social network in terms of accuracy metrics as well as user
\end{abstract}

%
%


\maketitle
Users need to find relevant friends in order to engage on any social network. Social platforms like Facebook, LinkedIn, Twitter facilitate friend discovery via {\em Friend Recommendation} \cite{dong2012link,backstrom2011supervised}. A good recommendation system strengthens the network by aiding creation of new social connections between existing users. It also helps in retention of new users by helping them find friends as they join the platform.
Hence, efficacy of friend recommendation method is of utmost importance to drive growth and engagement on the platform. 

The problem of friend recommendation fits into the classical link prediction problem \cite{liben2007link}. Given a snapshot of a social network at time $t$, can we accurately predict the edges that will be added to the network during the interval from time $t$ to a given future time $t'$? In a nutshell, can the current state of the network be used to predict future links? Traditional methods for link prediction were based on measures for analyzing the ``proximity'' of nodes in a network. Specifically, if the neighborhoods of two nodes have a large overlap, existing methods will indicate that they are highly likely to share a link. Common neighbors, Jaccard coefficient, Adamic-Adar \cite{Liben-Nowell:2007}, preferential attachment \cite{zeng2016link} are different measures that have been devised to assess overlap between node neighborhoods. Supervised models \cite{dong2012link} have also been trained with these features for link prediction.

 
Recently, there has been a lot of work in the development of methods for learning continuous representations \cite{graphrep,deepwalk,node2vec,tang2015line,cui2018survey,hong2017sena,hong2018page} of nodes that can effectively preserve their network neighborhoods. Such representations have proven to be more effective than hand-engineered features in encoding structural information for nodes in classification problems setup for link prediction \cite{deepwalk,node2vec}.

In this work, we present and evaluate a friend recommendation system for a real-world social network. We propose a framework for graph representation learning on heterogeneous networks with multiple edge types for link prediction. The system has two components - network embedding for a large heterogeneous network, and training a friend recommendation model  on a large set of known friend and non-friend pairs by leveraging the learned embedding. 

We split a heterogeneous network with multiple edge types into homogeneous components and obtain edge embedding for each component. Our friend recommendation system contains a multi-tower neural network which takes the homogeneous embeddings as inputs and combines them to obtain a unified edge embedding for the link prediction problem.   

Our contribution are as follows:
\begin{itemize}
\item We demonstrate the efficacy of network embedding for link prediction on a large real-world network.
\item We provide a formulation of friend recommendation problem as link prediction in an edge heterogeneous network.
\item We propose methods to obtain unified edge embedding by combining segregated embedding from homogeneous components. 
\item We present a multi-tower neural network architecture for learning unified edge embedding for the link prediction problem.
\item We evaluated the method by comparing it with the state-of-the-art approaches offline and also by deploying the system on an active platform.
\end{itemize}

\section{Network Embedding for Friend Recommendation}
We consider friend recommendation as a binary classification problem where a pair of users will be classified as {\em friends} or  {\em not-friends}. Given a collection of user-pairs (i.e., edges), we build a model that can learn to predict new edges. 

\subsection{Network Embedding}
Network embedding has shown great success in various social network applications, e.g., link prediction, node clustering, multi-label classification of nodes, etc. The idea is to learn a node-centric function that can map nodes into a low dimensional vector space by preserving the structural information about their neighborhoods. The node embedding of two nodes can be combined to form a representation of the edge connecting them. 
In case of link prediction, such an edge embedding can be given to a classifier to predict whether the edge is likely to exist or not. Two popular node embedding methods are as follows: 
\\
\\
{\bf DeepWalk} \cite{deepwalk} learns node embedding in a homogeneous network. Unbiased, uniform, fixed number of random walks of pre-decided length are generated starting at each node in the network to produce `sentences' of nodes, similar to sentences of words in a natural language. 
The Skip-gram algorithm devised by Mikolov et al. \cite{word2vec1} is used to obtain node embedding from the random walks, which are expected to capture the contextual properties of the network nodes -- the nodes that occur in same context have similar vector embedding.
\\
\\
\textbf{Node2vec} \cite{node2vec} is another method for homogeneous node embedding. A biased random walk to navigate the neighborhood of a node can be parameterized to make a transition from breadth-first search (BFS) to depth-first search (DFS). However, a proper parameterization is critical for good performance and this requires heavy tuning.

\subsection{Extending Network Embedding for Heterogeneous Multi-Graph}
Models like DeepWalk \cite{deepwalk} and Node2vec \cite{node2vec} are restricted to homogeneous networks. However, real-world social networks are heterogeneous in nature -- nodes are of multiple types such as users, posts, etc., and edges can be drawn based on different relationships such as friendship, like, comment, follow, etc. Existing network embedding methods designed for homogeneous networks may not be directly applicable to heterogeneous networks. Recently, metapath2vec \cite{metapath2vec}, an embedding technique for  heterogeneous networks was proposed, which defines {\em metapath} (a sequence of node types) to restrict the random walks. 
sHowever, it is not obvious how to define such a metapath as we often lack an intuition for the paths and metapaths, and the length of the metapath. Moreover for a heterogeneous network with different edge types and multiple edges between two nodes (heterogeneous multi-graph), there is no intuitive way to define a metapath of edges types. Chang et al. \cite{chang2015heterogeneous} propose a deep architecture to learn node embedding in a multi-modal network with image and text nodes. However, it doesn't generalize to a multi-graph where different types of edges exist between a pair of nodes. Next, we study straightforward extensions of DeepWalk for a heterogeneous multi-graph. 

\subsubsection{Equal Probability of Edges:}
Similar to DeepWalk, we generate unbiased random walks by assigning equal probability of walking through any edge between two nodes. 
This increases the probability of the random walk going to a node which has multiple edges from the current node. For instance, if two users $A$ and $B$ have each other listed as a contact, they chat and are also friends, then the random walker would be thrice as likely to traverse from $A$ to $B$ as compared to a setting where they shared a single edge. This technique suffers from the following limitation. For any node in the Hike network, there are far more contact edges than friends and even fewer chat connections. Thus the random walk has higher chances of going via contact edges, and in some cases completely avoid chat edges. This method will be referred as {\bf HeteroDeepWalk}, henceforth.

\subsubsection{Equal Probability of Edge Types:}
Here is a simple way to resolve the problem of  HeteroDeepwalk, i.e., random walks being biased by the dominant edge type. The random walk is generated in two steps: (i) an edge type is chosen randomly from all possible edge types, (ii) an edge is randomly chosen from all edges of the selected edge type. 
This amounts to biasing the random walks uniformly with equal weights for each edge type. This method will be referred as {\bf UniformBiasDeepWalk}, henceforth. In reality, different edge types contribute differently to the random walks and would have unequal weights. We don't have an intuitive way to obtain these weights.


\section{Proposed Solution: Heterogenous Edge Embedding}
\label{method}
HeteroDeepWalk and UniformBiasDeepWalk are two straightforward extensions of DeepWalk to deal with multi-graphs. However, the main drawback of these methods is that there is no obvious way to figure out how to bias the random walks for each edge-type; there is no reason to think that edge-types have equal importance. We propose a method to automatically estimate the weights of each edge-type in the embedding and obtain an edge embedding comprising of the contributions from various edge-types. Our method has four steps:
\begin{enumerate}
\small
\item Split the multi-graph into homogeneous sub-graphs each with one edge type.
\item Obtain node embedding from each of these subnetworks.
\item Obtain edge embedding from node embedding for each of these subnetworks.
\item Train a unified, heterogeneous edge embedding for link prediction  
\end{enumerate}
For example, we can split a social network into friend subnetwork, contact subnetwork and chat subnetwork, where users are connected via friendship, contact list and chatting, respectively. Each of these networks is homogeneous and can be embedded to a low dimensional space using DeepWalk  or Node2vec. Each node in the original heterogeneous network thereby has an embedding in three different spaces, e.g., for a node $v$, we have vector representations -- $v^{friend}$, $v^{contact}$, and $v^{chat}$. For simplicity, we assume the dimension of 3 spaces is equal.

Given these node representations, we can combine them in various ways to obtain the embedding of the edge $\langle u,v \rangle$ connecting two nodes $u$ and $v$. If $u^x$ is the node embedding of $u$ for the homogeneous x-subnetwork (where x = contact/chat/friend), then a homogeneous edge embedding $e_{\langle u,v \rangle}^x$, can be computed by taking average, Hadamard product or concatenation of $u^x$ and $v^x$. 

Next, we combine the segregated edge representations, $e_{\langle u,v \rangle}^x$, for different edges of type $x$ between a node pair, $\langle u,v \rangle$, to obtain a unified, heterogeneous edge embedding, $E_{\langle u,v \rangle}$. We discuss two methods based on neural network and logistic regression for doing this.\\

\begin{figure*}[t] 
  \begin{minipage}[b]{0.55\linewidth}
    \centering
    \includegraphics[width=.90\linewidth]{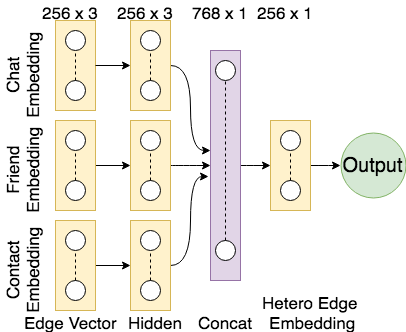} 
    \label{fig:axiom1}
  \end{minipage}
  \begin{minipage}[b]{0.50\linewidth}
    \centering
    \includegraphics[width=.90\linewidth]{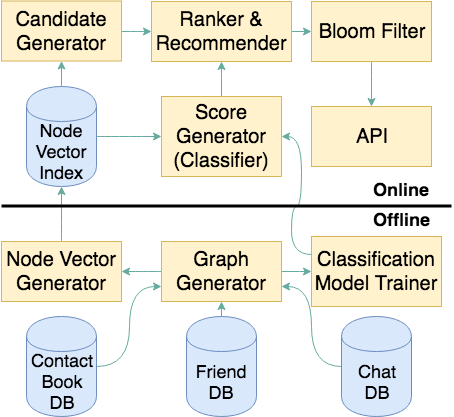}
    \label{fig:axiom2}
  \end{minipage} 
  \caption{(Left) Proposed neural network architecture (described in Section \ref{method}). (Right) System diagram (described in Section \ref{sec:system}). }\label{fig:model}
\end{figure*}

\noindent\textbf{Neural Network:}
Figure \ref{fig:model} (Left) shows the architecture of a neural network to train a heterogeneous edge embedding for link prediction. It takes the edge vectors from different sub-networks ({\em e.g.}, contact, friend and chat) as inputs. Each edge vector is passed through a hidden layer with 256 units that use RELU activation. The outputs are then concatenated and fed into another hidden layer that creates a 256 dimensional unified edge embedding. This unified embedding is passed into a Sigmoid layer to predict the class label - {\em link} or {\em no-link}.




\noindent\textbf{Logistic Regression:} We can learn a unified edge embedding as a linear combination of different dimensions present in the homogeneous edge representations, $e_{\langle u,v \rangle}^x$, for the set of edge types, $\{x\}$. The weights can be obtained by training a logistic regression model for link prediction that uses the edge vectors as features.
\section{System Description}\label{sec:system}

Delivering friend recommendations in a massive online social network within strict Service Level Agreements (SLAs) poses significant engineering challenges. Traditional recommendations, which utilized features to assess overlap in network neighborhood ({\em e.g.,} common neighbors) couldn't scale to deliver recommendations beyond 1 or 2 hops in the user's neighborhood. The reason being online computation of such features between a pair of nodes is quite expensive. Again, pre-computing such features is not an option for fast evolving social networks as they become stale soon. Since node embedding effectively captures network neighborhood for a node, it provides an elegant solution to this problem of finding recommendations that go beyond 1 or 2 hops. We discuss how this is achieved as we describe our system below.

Our system, as seen in Figure \ref{fig:model} (Right), has two parts: Offline and Online. The offline set up
recomputes the social graph daily, stores node embedding for each user in a scalable similarity search index \cite{faiss}, and trains a new recommendation model based on heterogeneous edge embedding as described in Section \ref{method}. In the online system, friend recommendations for a user are created in two steps. First, we look up the user's node embedding and perform a nearest neighbour search to fetch recommendation candidates. Second, we score these candidates using our neural network based model. A bloom filter check ensures that recommendations are not repeated for a user. Top $k$ recommendations generated in this manner are served in different recommendation widgets on the Hike app.

\section{Experimental Results}
\subsection{Dataset Description} The data for this work is taken from a subgraph of Hike network which contains users from a selected set of closely connected districts in one state. We divide the graph into two parts: pre-July containing the network structure on 30th June 2018, and edges added in the month of July. We employ the pre-July network comprising of 3.3 million nodes and 32 million edges for training network embedding. Another 10 million node-pairs (5 million edges and 5 million non-edges), which were not the part of the pre-July network, are used for training the friend recommendation model. Finally, the trained model was evaluated on a set of 1 million node-pairs, which also contains friendships made in July 2018 as positive examples. For training the embedding, we used a heterogeneous network 
with three edge types: `contact' if the users are in each other's contact book, `friend' if the users are friends, and `chat' if the users have chatted at least once. This setting allows for multiple edges between two nodes, e.g., if they are present in each other's contact list, friends and have chatted in July, there will be three edges between them. 

Our network is much larger than what has been previously studied in friend recommendation paradigm using network embedding \cite{deepwalk,node2vec}. The edge heterogeneity is also unique to our network which is a multi-network with multiple edge-types. Another confounding factor in our data is that 46\% of node-pairs in our validation data share no mutual friend -- there are many isolated and loosely connected nodes. These challenges compound the difficulty of the friend recommendation problem. 

\subsection{Comparative Evaluation} We compare different methods of computing edge embedding from node embedding for both node2vec and DeepWalk style random walks. Also, we evaluate the logistic regression model and the neural network model described in Section \ref{method}. The evaluation based on Area Under the ROC Curve (AUC) is performed on the link prediction problem cast as a classification of node pairs into `link' or `no-link'. To learn node embedding, we generate 10 random walks of length 30 emanating from each node. The context window size for the Skip-gram model \cite{fasttext} is taken to be 10, and we optimize the negative sampling loss with a learning rate of 0.01 to obtain embedding in 128-dimensional space. The results are presented in Table \ref{tab:EdgeCombiner2}. We find that the best performing edge embedding arises as a result of: (i) using `concatenate' on node2vec node-embeddings, (ii) unifying homogeneous edge vectors with the trained multi-tower neural network (Figure \ref{fig:model} (Left)). We refer to this variant as {\bf Node2Vec-NeuralNet}.

Further, we evaluate Node2Vec-NeuralNet, against the following baselines: (i) DeepWalk: learning node embedding using DeepWalk algorithm on the homogeneous friendship subnetwork , (ii) Node2vec: node embedding using node2vec on homogeneous friendship subgraph, (iii) HeteroDeepWalk, and (iv) UniformBiasDeepWalk.
The embeddings obtained using the above methods are fed to a logistic regression model for classification of node-pairs into `friend' and `non-friend'.
We employ three metrics for evaluation - AUC, Precision at 5 (P @ 5), and Real world Click-Through Rate (CTR) for top 1 recommendation.

The results are shown in Table \ref{tab:tab1}. Node2Vec-NeuralNet is significantly more accurate compared to other methods for predicting `friend' edges in our test dataset of 1 million node-pairs.  Node2Vec-NeuralNet beats the DeepWalk baseline by 20\% in AUC and 4.4\% in Precision @ 5.

We conducted controlled experiments on the Hike app to test the relevance of the top friend recommendation generated by each of the five model variants in the real world. We used a sample set of 10K users for each variant. The candidate set used to find recommendations for a user comprised of 1 hop neighbours of the user on the Hike network. In Table \ref{tab:tab1}, the last column compares the click through rates (CTR) for each variant to the DeepWalk baseline. Again, Node2Vec-NeuralNet shows a relative improvement of 7.61\% over DeepWalk.
\begin{table}
\scriptsize


\begin{tabular}{|c|c|c|c|}

  \hline
    Node & Edge  &  & \\ 
    Embedding & Combiner  & LogReg &  NeuralNet\\ 

\hline  
    & Average & 0.79 & 0.81 \\
    DeepWalk & Hadamard & 0.78 & 0.82 \\
    & Concatenate & 0.80 & 0.81 \\ \hline
    
    & Average & 0.79 & 0.82 \\
    Node2Vec& Hadamard & 0.78 & 0.81 \\
    & Concatenate & 0.80 & {\bf 0.84} \\
\hline
\end{tabular}

\caption{\scriptsize\strut AUC of diff. edge embedding}
\label{tab:EdgeCombiner2}


\begin{tabular}{|c|c|c|c|}

\hline

Model  & AUC & P @ 5 & \% Increase \\
& & & in CTR \\

\hline
DeepWalk & 0.70 & 0.91 & 0 \\
Node2Vec & 0.72 & 0.90 & 2.29 \\
HeteroDeepWalk & 0.71 & 0.83 & -0.67 \\
UniformBiasDeepWalk & 0.72 & 0.92 & 1.26 \\
Node2Vec-NeuralNet & {\bf 0.84} & {\bf 0.95} & {\bf 7.61} \\
\hline

\end{tabular}

\caption{\scriptsize\strut Offline and Real-World Evaluation}
\label{tab:tab1}


\end{table}
    


\section{Conclusion}
In this work, we developed and compared methods to learn edge embedding in a heterogeneous multi-graph. 
We showed the efficacy of such an edge embedding for deriving friend recommendation on Hike's network with a large scale offline evaluation as well as real world user experiments. 
Friend recommendation based on this method is deployed and currently running live at Hike.

%
%
\bibliographystyle{splncs04}

\begin{thebibliography}{10}
\providecommand{\url}[1]{\texttt{#1}}
\providecommand{\urlprefix}{URL }
\providecommand{\doi}[1]{https://doi.org/#1}

\bibitem{backstrom2011supervised}
Backstrom, L., Leskovec, J.: Supervised random walks: predicting and
  recommending links in social networks. In: Proceedings of the fourth ACM
  international conference on Web search and data mining. pp. 635--644. ACM
  (2011)

\bibitem{fasttext}
Bojanowski, P., Grave, E., Joulin, A., Mikolov, T.: Enriching word vectors with
  subword information. Transactions of the Association for Computational
  Linguistics  \textbf{5},  135--146 (2017)

\bibitem{chang2015heterogeneous}
Chang, S., Han, W., Tang, J., Qi, G.J., Aggarwal, C.C., Huang, T.S.:
  Heterogeneous network embedding via deep architectures. In: Proceedings of
  the 21th ACM SIGKDD International Conference on Knowledge Discovery and Data
  Mining. pp. 119--128. ACM (2015)

\bibitem{cui2018survey}
Cui, P., Wang, X., Pei, J., Zhu, W.: A survey on network embedding. IEEE
  Transactions on Knowledge and Data Engineering  (2018)

\bibitem{metapath2vec}
Dong, Y., Chawla, N.V., Swami, A.: Metapath2vec: Scalable representation
  learning for heterogeneous networks. In: Proceedings of the 23rd ACM SIGKDD
  International Conference on Knowledge Discovery and Data Mining. pp.
  135--144. KDD '17, ACM, New York, NY, USA (2017)

\bibitem{dong2012link}
Dong, Y., Tang, J., Wu, S., Tian, J., Chawla, N.V., Rao, J., Cao, H.: Link
  prediction and recommendation across heterogeneous social networks. In: 2012
  IEEE 12th International Conference on Data Mining. pp. 181--190. IEEE (2012)

\bibitem{node2vec}
Grover, A., Leskovec, J.: node2vec: Scalable feature learning for networks.
  CoRR  \textbf{abs/1607.00653} (2016), \url{http://arxiv.org/abs/1607.00653}

\bibitem{graphrep}
Hamilton, W.L., Ying, R., Leskovec, J.: Representation learning on graphs:
  Methods and applications. CoRR  \textbf{abs/1709.05584} (2017),
  \url{http://arxiv.org/abs/1709.05584}

\bibitem{hong2017sena}
Hong, S., Chakraborty, T., Ahn, S., Husari, G., Park, N.: Sena: preserving
  social structure for network embedding. In: Proceedings of the 28th ACM
  Conference on Hypertext and Social Media. pp. 235--244. ACM (2017)

\bibitem{hong2018page}
Hong, S., Park, N., Chakraborty, T., Kang, H., Kwon, S.: Page: Answering graph
  pattern queries via knowledge graph embedding. In: International Conference
  on Big Data. pp. 87--99. Springer (2018)

\bibitem{faiss}
Johnson, J., Douze, M., J{\'e}gou, H.: Billion-scale similarity search with
  gpus. arXiv preprint arXiv:1702.08734  (2017)

\bibitem{liben2007link}
Liben-Nowell, D., Kleinberg, J.: The link-prediction problem for social
  networks. Journal of the American society for information science and
  technology  \textbf{58}(7),  1019--1031 (2007)

\bibitem{Liben-Nowell:2007}
Liben-Nowell, D., Kleinberg, J.: The link-prediction problem for social
  networks. J. Am. Soc. Inf. Sci. Technol.  \textbf{58}(7),  1019--1031 (May
  2007)

\bibitem{word2vec1}
Mikolov, T., Chen, K., Corrado, G., Dean, J.: Efficient estimation of word
  representations in vector space. CoRR  \textbf{abs/1301.3781} (2013),
  \url{http://arxiv.org/abs/1301.3781}

\bibitem{deepwalk}
Perozzi, B., Al-Rfou, R., Skiena, S.: Deepwalk: Online learning of social
  representations. In: Proceedings of the 20th ACM SIGKDD International
  Conference on Knowledge Discovery and Data Mining. pp. 701--710. KDD '14,
  ACM, New York, NY, USA (2014)

\bibitem{tang2015line}
Tang, J., Qu, M., Wang, M., Zhang, M., Yan, J., Mei, Q.: Line: Large-scale
  information network embedding. In: Proceedings of the 24th International
  Conference on World Wide Web. pp. 1067--1077. International World Wide Web
  Conferences Steering Committee (2015)

\bibitem{zeng2016link}
Zeng, S.: Link prediction based on local information considering preferential
  attachment. Physica A: Statistical Mechanics and its Applications
  \textbf{443},  537--542 (2016)

\end{thebibliography}

\end{document}